\documentstyle[12pt]{article}
\author{S. V. Krasnikov\thanks{Email: \it redish@pulkovo.spb.su}\\
The Central\\
Astronomical Observatory at Pulkovo, St Petersburg,\\
196140, Russia}
\title{On the quantum stability of the time machine}
\newcommand{\ssy}[5]{#1 #4 {\it #2} {\bf #3} #5}

\begin{document}
\maketitle
\begin{abstract}
In a number of papers it has been claimed that the time machine are quantum
unstable, which manifests itself in the divergence of the vacuum expectation
value of the stress-energy tensor $\langle{\bf T}\rangle$
near the Cauchy horizon. The expression for $\langle{\bf T}\rangle$ was found
in these papers on the basis of some specific approach \cite{Fro,KimT}.\par
We show that this approach is untenable in that the above expression
firstly is not derived from some more fundamental and undeniable premises,
as it is claimed, but rather postulated and secondly contains
undefined terms, so that one
can neither use nor check it. As a counterexample we cite a few cases of
(two-dimensional) spacetimes containing time machines with $\langle{\bf T}%
\rangle$ bounded near the Cauchy horizon.\par
\end{abstract}
\section{Introduction}
Since the wormhole-based time machine was proposed \cite{Tho}
much efforts have been directed towards finding a mechanism that could
"protect causality" and destroy such a time machine. One of the most
popular ideas is that the creation of
the time machine might be prevented by quantum effects since as it is
claimed in \cite{KimT} "at any event in spacetime, which can be joined to
itself by a closed null geodesic, the vacuum fluctuations of a massless
scalar field should produce a divergent renormalized stress-energy
tensor". The considerations leading to such a claim I shall call hereafter
"FKT approach". \par

In essence, the FKT approach amounts to the following \cite{Fro,KimT} (see
also \cite{Kli}). The vacuum
expectation value of the stress-energy tensor
$\langle T_{\mu\nu}\rangle$ of the field $\phi$ in the (multiply
connected) spacetime $M$ containing a time machine is found by
applying some
differential operator $D_{\mu\nu}$ to the Hadamard function
\begin{equation}
\label{DefG}
G^{(1)}(X,X')\equiv\bigl\langle\{\phi(X),\phi(X')\}\bigr\rangle.
\end{equation}
To find $G^{(1)}$ it is proposed to use the formula
 \begin{equation}
\label{Cov}
G^{(1)}(X,X^{\prime})=G^\Sigma\equiv
\sum_n{\widetilde G}^{(1)}(X, \gamma^n X^{\prime}) .
\end{equation}
Here $\widetilde G^{(1)}$ is the Hadamard function of $\phi$ in the
spacetime $\widetilde M$, which is the universal covering space for $M$,
and $\gamma^n X\in \widetilde M$ is the $n$-th inverse image of $X\in M$
($\gamma^0X$ is identified in (\ref{Cov}) with $X$). The advantage of the
use of (\ref{Cov}) is that $\widetilde G^{(1)}$ is supposed to have the
Hadamard form:
\begin{equation}
\label{Had}
{\widetilde G}^{(1)}(X,X')=\tilde u\sigma^{-1}+\tilde v\ln{|\sigma|}
+\mbox{nonsingular terms}
\end{equation}
where $\sigma$ is half the square of the geodesic distance between $X$ and
$ X'$, and $\tilde u,\ \tilde v$ are some smooth functions. We might think
thus that
\begin{equation}
\begin{array}{lcl}\displaystyle{
\langle T_{\mu\nu}\rangle^{{\rm ren}}_{M\vphantom{\widetilde M}}
                               }&=&
\displaystyle{\langle T_{\mu\nu}\rangle^{{\rm ren}}_{\widetilde M} + \sum
_{n\not=0}\lim_
{ X'\to X\atop{}} D_{\mu\nu}{\widetilde G}^{(1)}
(X, \gamma^n X^{\prime})}\\
\strut&&\\&\to&
\displaystyle{
\langle T_{\mu\nu}\rangle^{{\rm ren}}_{\widetilde M} + \sum
_{n\not=0}\;\lim_
{X'\to X\atop X\to\, \mbox{\scriptsize horizon}} D_{\mu\nu}
(\tilde u\sigma_n^{-1}+\tilde v\ln{|\sigma_n|}).
                              }\end{array}
\label{Tren}
\end{equation}
Here $\sigma_n\equiv \sigma(X,\gamma^nX')$ and the subscript "ren"
(renormalized) has appeared  because renormalization of
$\langle {\bf T}\rangle_{\widetilde M}$ and of
$\langle {\bf T}\rangle_{M\vphantom{\widetilde M}}$
requires substraction of the same terms. The last series in (\ref{Tren})
diverges (since $\sigma_n\to0$, when $X$ approaches the horizon), so the
conclusion is made that the appearance of a closed
timelike curve must be prevented (unless some effects of quantum gravity
remedy the situation) by the infinite increase of the energy density.
\smallskip \par

The goal of this paper is to state that there is actually {\em no\/} reasons
to expect that the energy density diverges at the Cauchy horizon in the
general case. In particular, we cite a few examples in which the
stress-energy tensor of the massless scalar field (as found in the usual way
--- through quantization directly in $M$) remains bounded as the horizon
is approached. These examples are not of course anywhere near an adequate
model of the time machine (for example, the wave equation on the
two-dimensional cylinder has no solutions, except for constant, continuous
at the Cauchy horizon). However, they can well serve as counterexamples to
statements like that cited above. So it seems important to find out
whether they indicate only some loop-holes in the FKT approach, which can
then be used in the general case, or whether this approach must be fully
revised. Our analysis in Section~2 shows that the latter is true --- by
several independent reasons one cannot extract any information from expression
(\ref{Tren}). In Section~3 some special cases are considered necessary to
support statements from  Section~2.
\section{Analysis}


\subsection{Going to the universal covering}
Formula (\ref{Cov}), combined with some implicit assumptions serves as a
basis for
the overall FKT approach since one cannot use (\ref{Had}) in multiply
connected spacetimes, where $\sigma$ is not defined. To discuss its validity
and to reveal these assumptions
let us first state the simple fact that most properties of the Hadamard
function, {\em including\/} the validity of (\ref{Had}), depends on the
choice of the vacuum appearing in  definition (\ref{DefG}). So,
formulas like (\ref{Cov}) are meaningless until we specify the vacuua
$|0\rangle $ and $|\tilde 0\rangle $ in $M$ and ${\widetilde M}$
respectively. We come thus to the problem of great importance in our
consideration --- how, given $|0\rangle $, one could determine corresponding
$|\tilde 0\rangle $?
The above-mentioned assumptions concern just this problem. They must
be something like the following:
\begin{enumerate}
\item  For any vacuum on $M$ there exists a vacuum $|\tilde 0\rangle $ on $
\widetilde{M}$ such that (\ref{Cov}) holds.
\item  The function $\widetilde{G}^{(1)}$ corresponding to $|\tilde 0\rangle $
has the "Hadamard form" (\ref{Had}).
\item  ${G}^{(1)}$ determines $\widetilde{G}^{(1)}$ uniquely.
\end{enumerate}\par

The validity of Assumption 1 is almost obvious in the simplest cases (see
below), but it was not proven in the general
case. (One can meet the references to \cite{Ban} in this connection. Note,
however, that the functions $K_C$ which stand there in the analog of our
formula (\ref{Cov}), are actually not defined%
\footnote{I am
grateful to Dr. Parfyonov, who explained to me this issue.} in our case,
i.~e.\ when $|\Gamma|=\infty$.)

Assumption 2 seems still more arbitrary. The validity of (\ref{Had}) was
proven not for {\em any\/} state, but only for some specific class of states
(see \cite[Sect.~2c]{Ful}) and there is no reason to believe that our $%
|\tilde0\rangle$ belongs just to this class.

Assumption 3 is definitely untrue. In the following section we construct
as an example a
class of vacuua $|\tilde0\rangle_f$ such that (\ref{Cov}) is satisfied for
any $f$ while $\widetilde{G}_f$ differ for different $f$. This
nonuniqueness is far from harmless. As we argue below it makes, in fact,
expression (\ref{Tren}) meaningless.
\subsection{The expression  for the stress-energy tensor}
%
Expression (\ref{Tren}) is the main result of the FKT approach
and (\ref{Cov}) is needed only to justify it. So let us state first that
\begin{enumerate}
\item  (\ref{Tren}) does not follow (or, at least, does not follow
immediately) from (\ref{Cov}), since
\begin{enumerate}
\item To write $\lim D_{\mu \nu }\sum \widetilde G^{(1)}=\sum \lim D_{\mu
\nu } \widetilde G^{(1)}$ without a special proof one must be sure that
the series
$\sum \widetilde G^{(1)}$ and $\sum D_{\mu \nu }\widetilde G^{(1)}$ converge
uniformly, while it is clear that they do not (at least as long as (\ref{Had}%
) holds). This nonuniformity manifests itself, in particular, in the fact
that, in general, one cannot drop the nonsingular terms in
$\widetilde G^{(1)}$. In Subsection~$3.2$ we shall show that the last
series in (\ref{Tren}) can diverge {\em off\/} the Cauchy horizon though
(\ref{Had},\ref{Cov}) hold and $G^{(1){\rm ren}}$ (and $\langle {\bf
T}\rangle^{{\rm ren}}$) are smooth there.
\item  Even when $|\tilde 0\rangle $ belongs to the above-mentioned class, (%
\ref{Had}) is proven not for {\em any\/} $X,\ X^{\prime }$, but only for $%
X^{\prime }$ lying in the ''sufficiently small'' neighborhood of $X$. It is
necessary, in particular, that $\sigma (X,X^{\prime })$ would be defined
uniquely. To provide this in Ref. \cite{Ful}, for example, $X$ and $%
X^{\prime }$ are required not to lie respectively near points
$\underline{\vphantom{y}x},
\mbox{ and }\underline{y}$  connected by a null geodesic with a
point conjugate to $\underline{\vphantom{y}x}$ before $\underline{y}$. To
violate this condition for the points $X'$ and $\gamma X'$ it suffices to
separate the mouths of the wormhole widely enough and to fill the space
between them with the conventional matter \cite[Prop. 4.4.5]{HawEl}.
\end{enumerate}
\vbox{Thus, we see that (\ref{Tren}) must be regarded as an
independent assumption. We can, however, neither use nor check it in view
of the aforementioned ambiguity:}
\item Like the Hadamard function,
$\langle T_{\mu\nu}\rangle^{{\rm ren}}_{\widetilde M}$
depends on which vacuum we choose, while from the FKT standpoint all
vacuua $|\tilde0\rangle$ satisfying (\ref{Cov}) are equivalent. This
equivalence is of a fundamental nature --- the only physical object is the
spacetime $M$, while $\widetilde M$ and $|\tilde0\rangle$ are some
auxiliary matters and as long as (\ref{Cov}) holds we cannot apply any
extraneous criteria to distinguish among them. So, we have no way
 of determining what to substitute in (\ref{Tren}) as
$\langle T_{\mu\nu}\rangle^{{\rm ren}}_{\widetilde M}$.
In Subsection~$3.2$ we shall see that choosing different $|\tilde0\rangle$
(even when $\widetilde M$ is a part of the Minkowski plane)
one can make $\langle T_{\mu\nu}\rangle^{{\rm ren}}_{\widetilde M}$ finite
or infinite at the horizon at will.\par
\end{enumerate}
Let me note in passing that there is no point in using (\ref{Tren}) unless
we decide that $|\tilde0\rangle$ is among the very "good" and convenient
vacuua. For an arbitrary $|\tilde0\rangle$,
 it is not a bit easier to find $\langle {\bf T}\rangle^{{\rm
ren}}_{\widetilde M}$ than  $\langle {\bf T}\rangle^{{\rm
ren}}_{M\vphantom{\widetilde M}}$.
\subsection{Interpretation}
Suppose that $\langle\Psi|{\bf T}|\Psi\rangle^{{\rm ren}}_M$ for some
 $|\Psi\rangle $ does diverge
at the Cauchy horizon. Suppose further that it is $\langle {\bf
T}\rangle^{{\rm ren}}_M$ that stands in the right side of the Einstein
equations (though it is not obvious, see \cite{Bir} for the literature and
discussion). Does this really mean that owing to the quantum effects the
time machine $M$ cannot be created? I think that the answer is negative.
It well may be that
$\langle\Phi|{\bf
T}|\Phi\rangle^{{\rm ren}}_M$ does not diverge for some other state $
|\Phi\rangle$ (example see in
Subsection~$3.3$).
Why must we restrict ourselves to the state $|\Psi\rangle$? To prove that
the Einstein equations and QFT are incompatible in $M$ one must have
proven that the expected stress-energy tensor tends to
infinity for {\em any\/} quantum state, or at least for any state satisfying
some reasonable physical conditions (say, stability).

%

\section{Examples}
Let us find the expectation value of the stress-energy tensor in a few
specific cases. We restrict our consideration to the two-dimensional
cylinder $M$ obtained from the plane $(\tau ,\chi )$ by identifying
$\chi\leadsto \chi +H $ and endowed with the metric
\begin{equation}
\label{Metr}
ds^2=C(-d\tau^2+d\chi^2)=C\,dudv .
\end{equation}
Here $u\equiv\chi-\tau,\ v\equiv\chi+\tau$; $C$ is a smooth function on $M$.
To find in the ordinary way $\langle {\bf T}\rangle $ for the free real
scalar field $\phi $
\begin{equation}
\label{Weq}\Box \phi =0,\qquad \phi (\chi +H,\tau )=
\cases{\phi(\chi,\tau)&for the non-twisted field\cr
-\phi(\chi,\tau)&for the twisted field\cr}
\end{equation}
we must first of all specify the vacuum we consider. That is we must choose
a linear space of solutions of (\ref{Weq}) and an "orthonormal" basis \cite
{Bir} $U=\{u_n\}$ in it. In particular, this will define the Hadamard
function:
$$
G^{(1)}(X,X')=\sum_n u_n(X)u^*_n(X')+\mbox{complex conjugate}.
$$
A possible choice of $U$ for the non-twisted field is
\begin{equation}
u_n=|4\pi n|^{-1/2}e^{2\pi i H^{-1}(n\chi -|n|\tau )} \quad
n=\pm1,\pm2\dots \label{ConVac}
\end{equation}
The vacuum $|0\rangle_C$ defined by (\ref{ConVac}) (the "conformal" vacuum) is
especially attractive as the expressions for the Hadamard function $G_C^{(1)}$
and for the stress-energy tensor $\langle {\bf T}\rangle _C $ are already
obtained (see \cite[the neighborhood of formula (6.211)]{Bir}):
\begin{eqnarray}
\nonumber\langle T_{ww}\rangle _C^{\rm ren} &=&
-\frac{\pi\epsilon}{12H^2}+\frac{1}{24\pi}
\left[
\frac{C,_{ww}}{C}-\frac{3}{2}\frac{{C,_w}^2}{C^2}\label{Tcon}
\right]
,\quad w=u,v\\
&&\strut\\
\nonumber\langle T_{uv}\rangle _C^{\rm ren}&=&
\langle T_{vu}\rangle _C^{\rm ren}=-RC/(96\pi).
\end{eqnarray}
Here $\epsilon=-1/2$ or 1 depending on whether $\phi$ is twisted
or untwisted and $R$ is the curvature of $M$.
Though the absence of a solution corresponding to $n=0$ in (\ref{ConVac})
may seem artificial, it is, in
fact, an inherent feature of $|0\rangle_C$, which is to describe the vacuum
of $\phi$ as a massless limit of the "natural" vacuum of a massive field (cf
\cite[below (4.220)]{Bir}). One could start, however, from another vacuum
for the massive field  and arrive at another theory (see below) with the
basis $U^{\prime}$
$$
U^{\prime}=
U\cup u_0\equiv
(2H)^{-1/2}(F\tau+i/F) .
$$
Where the real constant $F$ is a free parameter. Choosing different
$F\neq0$ we
obtain different vacuua $|0\rangle_F$ and Hadamard functions $G^{(1)}_F$.
It is easy to see that
\begin{equation}
\label{GF}G_F^{(1)}=G_C^{(1)}+\frac{F^2}{H}\tau\tau^{\prime}+const.
\end{equation}
%
\subsection{Two-dimensional time machines in the conformal vacuum state}
As a first example let us consider the Misner spacetime, which is the
 quadrant $\alpha <0,$ $\beta >0$ of the  Minkowski plane $%
ds^2=d\alpha d\beta $ with points identified by the rule $(\alpha _{0,}\beta
_0)\mapsto (A\alpha _{0,}\beta _0/A)$. The coordinate transformation%
$$
u=-W^{-1}\ln |W\alpha |,\quad v=W^{-1}\ln (W\beta )
$$
delivers the isometry between Misner space and $M$ with%
$$
C=e^{2W\tau },\qquad H=W^{-1}\ln A .
$$
$W$ here is an arbitrary parameter with dimension of mass. Substituting this
in (\ref{Tcon}) we immediately find%
$$
\langle T_{ww'}\rangle_C^{\rm ren}=
-W^2\left(\frac{\epsilon\pi}{12\ln^2A}+\frac{1}{48\pi}\right)\delta_{ww'}.
$$
\noindent The metric in coordinates $\alpha,\beta$ is "good" (smooth,
nondegenerate) near the Cauchy horizons $\alpha=0$ or $\beta=0$. So, the
proper basis of an observer approaching to one of them with a finite
acceleration
is related to the basis $D\equiv\{\partial_\alpha,\partial_\beta\}$ by a
finite Lorentz transformation. Thus the quantities we are to examine are,
in fact, the components of $\langle {\bf T}\rangle_C^{\rm ren}$ in the
basis $D$, which are
$$
\begin{array}{c}
\langle T_{\alpha\alpha}\rangle_C^{\rm ren}=T\alpha^{-2},\quad
\langle T_{\beta\beta}\rangle_C^{\rm ren}=T\beta^{-2},\quad
\langle T_{\alpha\beta}\rangle_C^{\rm ren}=
\langle T_{\beta\alpha}\rangle_C^{\rm ren}=0,
\\ \strut\\
\displaystyle T\equiv -\left(\frac{\pi\epsilon}{12\ln^2A}+
\frac{1}{48\pi}\right).
\end{array}
$$\par

Now let us use the above simple method to find $\langle {\bf
T}\rangle^{\rm ren}$ for two time machines more (see also \cite{Yur}).
Consider first the cylinder $S$ obtained from the strip 
\begin{equation}
\begin{array}{c}
ds^2=W^{-2}\xi^{-2}(-d\eta^2+d\xi^2)=\xi^{-2}d\alpha d\beta, \\
\strut \\ \mbox{where }\alpha\equiv(\xi-\eta)/W,\
\beta\equiv(\xi+\eta)/W;\quad
\eta\in(-\infty,\infty),\;\xi\in[1,A]. \label{StMod}
\end{array}
\end{equation}
by gluing points $\eta=\eta_0,\ \xi=1$ with the points $\eta=A\eta_0,\ \xi=A$.
This spacetime was considered in detail in \cite{Fro} where it was called
the "standard model". A simple investigation shows that the Cauchy
horizons $\alpha=0$ and $\beta=0$ divide $S$ into three regions. Causality
holds in the "inner" region $\widetilde S:\ \alpha, \beta>0$ and violates
in $I^\pm(\widetilde S)$. Introducing new coordinates $u,\;v$:
$$
u\equiv W^{-1}\ln \alpha,\quad v\equiv W^{-1}\ln \beta
$$
we find that $\widetilde S$ like the Misner space%
\footnote{In spite of their apparent similarity these spaces are 
 significantly distinct. For example, the Misner spacetime is geodesically
incomplete \cite{HawEl}, and the standard model is not \cite{Me}. This
may be of importance if one would like to separate $X$ and $X'$ "widely
enough" (see item 1b in the previous section).}
 is isometric to $M$.
This time
$$
C=\cosh^{-2}W\tau,\quad H=W^{-1}\ln A,
$$
which yields
\begin{equation}
\begin{array}{c}
\langle T_{\alpha\alpha}\rangle_C^{\rm ren}=T\alpha^{-2},\quad
\langle T_{\beta\beta}\rangle_C^{\rm ren}=T\beta^{-2},\\
\strut\\
\langle T_{\alpha\beta}\rangle_C^{\rm ren}=
\langle T_{\beta\alpha}\rangle_C^{\rm ren}=(1/12\pi)(\alpha+\beta)^{-2}.
\label{TFro}
\end{array}
\end{equation}\par

Consider lastly the spacetime obtained by changing $\xi^{-2}\to\eta^{-2}$
in (\ref{StMod}). This spacetime is similar to the standard model, but has
a somewhat more curios causal structure --- there are two causally
nonconnected regions separated by the time machine. $\langle {\bf
T}\rangle_C^{\rm ren}$ differs from that in (\ref{TFro}) by the
off-diagonal (bounded) terms
$$
\langle T_{\alpha\beta}\rangle_C^{\rm ren}=
\langle T_{\beta\alpha}\rangle_C^{\rm ren}=-(1/12\pi)(\alpha-\beta)^{-2}.
$$\par

So, we see that in all three cases the vacuum energy density (associated
with  {\em some} vacuum states)
does grow infinitely as one approaches to the Cauchy horizon. A few
comments are necessary, however:
%
%
\begin{enumerate}
\item
The divergence in discussion is not at all something peculiar to the time
machine: the passage to the limit $A\to\infty$ shows that
precisely
the same divergence (with $T=-1/(48\pi)$) takes place in $\widetilde M$ though
(in the case of Misner space) $\widetilde M$ is merely
a part of the Minkowski plane. This suggests that for the time machine
too, the divergence of the stress-energy tensor is a consequence not of
its causal or topological structure but rather of the unfortunate choice
of the quantum state.
\item The twisted field at $A=e^{\sqrt{2}\pi}$  has the
bounded $\langle {\bf T}\rangle_C^{\rm ren}$ (cf. \cite{Sush}).
\item Let us consider  nonvacuum states now (see Subsection~$2.3$). The
first example is a two-particle state $|1_n1_{-n}\rangle$ with the particles
corresponding to the $n$-th and $-n$-th modes of (\ref{ConVac}).
$\langle1_{-n}1_n| {\bf T}|1_n1_{-n}\rangle^{\rm ren}$ is readily found using
\cite[eq.~(2.44)]{Bir}:
$$
\langle1_{-n}1_n| T_{\gamma\gamma}|1_n1_{-n}\rangle^{\rm
ren}=T'\gamma^{-2},\quad
\langle1_{-n}1_n| T_{\alpha\beta}|1_n1_{-n}\rangle^{\rm ren}
=\langle T_{\alpha\beta}\rangle^{\rm ren}_C
$$
with $T'\equiv T+2\pi nH^{-2},$ and $ \gamma\equiv\alpha,\beta$. Thus we
see that there {\em are\/}  states
with the bounded energy density of the untwisted field.\par

Another yet example is the equilibrium state at a nonzero temperature
$|t\rangle$. Expression (4.27) of  \cite{Bir} gives
$$
\langle t| T_{ww}| t\rangle^{\rm ren}
=\langle T_{ww}\rangle^{\rm ren}_C+\frac{\pi}{2H^2}\sum_{m=1}^\infty
\sinh^{-2}\frac{\pi m}{k_BtH}.
$$
So, for any $H$  there exists such temperature $t$ that
$\langle t| T_{\gamma\gamma}| t\rangle^{\rm ren}$ does not diverge at the
horizon.
\end{enumerate}
\subsection{Another vacuum}
%
The conformal vanuum is not suited for verifying or exemplifying most of
statements made in Section 2., since the Hadamard function does not exist
in this state. So consider now the new vacuum $|0\rangle_f$ on the plane
$(\tau,\chi)$
defined by the modes
\begin{equation}
u'_p\equiv\cases{
\displaystyle\frac{1}{ 2\sqrt{\pi\omega} } e^{ip\chi-i\omega\tau},&
$\omega\geq\delta$\cr
\strut&\cr
\displaystyle\frac{1}{ 2\sqrt{\pi} } e^{ip\chi}
(f^{-1}\cos\omega\tau-i\omega^{-1}f\sin\omega\tau),&
$\omega<\delta$.\cr}
\label{AnVac}
\end{equation}
where $\omega\equiv|p|$, $\delta$ is an arbitrary positive constant:
$\delta<1$ and $f$ is an arbitrary smooth positive function:
$f(\omega\geq \delta)=\sqrt\omega$. The modes (\ref{AnVac}) are obtained
>from that
defining the conformal vacuum on the plane by a Bogolubov transformation of
the low-frequency modes so as to avoid the infrared divergence without
affecting the ultraviolet behavior of $\langle {\bf T}\rangle$.
The asymptotic form of ${\widetilde G}^{(1)}_f$ does not depend  on $f$:
\begin{equation}
\forall f\qquad \widetilde G^{(1)}_f=-1/(2\pi)\ln|\Delta u\Delta
v|+\mbox{smooth, bounded function} .
\label{GAs}
\end{equation}
If we retain only the first term, we obtain (in the flat case)
$$
\lim_{X^{\prime}\to X} D_{\alpha\alpha}{\widetilde G}^{(1)}(X,\gamma^nX')=
\ln^2(A)\, \alpha^{-2}A^{-n}n^{-2}.
$$
So, the last series in (\ref{Tren}) diverges not only at the horizon, but
everywhere on $M$ (cf. Subsection $2.2$).\par

$\langle{\bf T}\rangle_f$ can be found from (\ref{GF}) (see
\cite[Sect.~$6.4$]{Bir}). For any $C$  we have:
$$
\langle T_{ww}\rangle_f^{\rm ren}=
\frac{1}{8\pi}\left[
-\frac{\;W^2}{6}+  \int^\delta_0
(f^{-2}\omega^2 +f^2-\omega)
\,d\omega\right],\quad
\langle T_{uv}\rangle_f^{\rm ren}=
\langle T_{uv}\rangle_C^{\rm ren}.
$$
Having taken an appropriate $f(\omega)$ one can make $\langle
T_{\alpha\alpha}\rangle^{\rm ren}_f$ infinite or zero at the horizon, as
we have stated in Subsection $2.2$.\par

To illustrate some more statements from Section~2.\ let
us first find $G^\Sigma$. To this end note that it has the form
\begin{equation}
G^\Sigma=\sum_n\int_{-\infty}^{\infty}h(p)e^{inHp}\,dp + c.\ c.
\label{GSum}
\end{equation}
with
$$
h\equiv\cases{
\displaystyle\frac{1}{ 4\pi\omega } e^{ip\Delta\chi-i\omega\Delta\tau},&
$\omega\geq1/2$\cr
\strut&\cr
\displaystyle\frac{1}{ 4\pi } e^{ip\Delta\chi}
(f^{-2}\cos\omega\tau\cos\omega\tau'+
\omega^{-2}f^2\sin\omega\tau\sin\omega\tau')
,&
$\omega<1/2$.\cr}
$$
The function $h(p)$ can be written as a sum: $h=(h-h_0)+h_0$, where
$$
h_0\equiv\frac{1}{4\pi\sqrt{1+p^2}}\;e^{ip\Delta\chi}
\; e^{-i\sqrt{1+p^2}\Delta\tau}.
$$
The first summand is a smooth function falling off at infinity like
$p^{-2}$ and
the second summand ($h_0$) is a holomorphic (but for $p=\pm i$) function
admitting the following estimate:
$$
|h_0|\leq C|x|^{-1/2}\;e^{|(\Delta\chi-\Delta\tau)y|}.
$$
Hence \cite{Fed} we can apply the Poisson formula to (\ref{GSum})
and obtain:
$$
G^\Sigma=2\pi H^{-1}\sum_n h(2\pi H^{-1}n) +c.\ c.
$$
We see thus that $G^\Sigma$ is indeed the Hadamard function and it
corresponds to the vacuum
$|0\rangle_F$ with $F=f(0)$.\par

{\em Remark 1.\/} This does not mean, however, that $G^\Sigma$ will be a
Hadamard function of some reasonable state for {\em any\/} $\widetilde
G^{(1)}$.
One can easily construct, for example, such a vacuum that $\widetilde
G^{(1)}(\chi,\chi')$ will not be invariant under translations
$\ \chi,\chi'\mapsto\chi+H,\chi'+H$ and
$G^\Sigma(\chi,\chi')$, as a consequence, will not even be symmetric.\par

{\em Remark 2.\/} For all $G^{(1)}_f$ with the same $f(0)$ the Hadamard
functions
$G^\Sigma$ are the same. This proves our statement from Subsection~$2.1$.

To find $\langle {\bf T}\rangle_F$ note that it differs from
$\langle {\bf T}\rangle_C$ only by the term arising from the second
summand in (\ref{GF})
(cf.~\cite[ eqs. (4.20), (6.136)]{Bir}):
$$
\Delta\langle T_{ww'}\rangle^{\rm ren}=\frac{F^2}{2H}
(\tau,_w\tau,_{w'}-
1/2\,\eta_{ww'}\eta^{\lambda\delta}\tau,_\lambda\tau,_\delta).
$$
That is
\begin{eqnarray*}
\langle T_{ww}\rangle_F^{\rm ren}&=&
\displaystyle{\frac{F^2}{8H\vphantom{\ln^2A}}-
W^2\left(\frac{\epsilon\pi}{12\ln^2A}+
\frac{1}{48\pi}\right)},\\
&&\strut\\
\langle T_{uv}\rangle_F^{\rm ren}&= &
\langle T_{vu}\rangle_F^{\rm ren}=
\langle T_{uv}\rangle_C^{\rm ren} .
\end{eqnarray*}
So, for all three time machines considered here there exists a vacuum,
such that the expectation value of the stress-energy tensor is bounded in
the causal region.
\section{Conclusion}
Thus, we have seen that one cannot obtain any information about the energy
density near the Cauchy horizon employing the FKT approach. In the absence
of any other general approach this means that all we have is a few simple
examples. In some of them the energy
density diverges there and in some do not. So, the time machine perhaps is
stable and perhaps is not. This seems to be the most strong assertion we
can make.

\end{document}